\documentclass[pre,superscriptaddress,showpacs,reprint]{revtex4-1}
\usepackage[latin1]{inputenc}
\usepackage[dvips]{graphicx}
\usepackage{verbatim}
\usepackage{amsmath}
\usepackage{amsfonts}
\usepackage{amssymb}
\usepackage{latexsym}

\newcommand{\beq}{\begin{equation}}
\newcommand{\eeq}{\end{equation}}
\newcommand{\pr}{\sigma}
\newcommand{\Spa}{S}

\newcommand{\pro}{p}
\DeclareMathOperator*{\argmax}{arg\,max}
\newcommand{\cass}{C_{ass}}
\newcommand{\NULL}{0}
\newcommand{\N}{t}

\newcommand{\hv}{{feature}}
\newcommand{\hvs}{{features}}
\newcommand{\hs}{{feature space}}

\begin{document}

\title{Features and heterogeneities in growing network models} 

\newcommand{\eqcontr}{\thanks{These authors contributed equally to the work.}}

\author{Luca Ferretti} 
\eqcontr
\affiliation{Centre de Recerca en AgriGen\`omica, Universitat Aut\`onoma de Barcelona, 08193 Bellaterra, Spain}
\email{Email: luca.ferretti@gmail.com}

\author{Michele Cortelezzi}
\eqcontr
\affiliation{Dipartimento di Fisica, Universit\`a di Pisa, Largo Bruno Pontecorvo 3, 56127 Pisa, Italy}

\author{Bin Yang}
\affiliation{Centre de Recerca en AgriGen\`omica, Universitat Aut\`onoma de Barcelona, 08193 Bellaterra, Spain}
\affiliation{Department of Food and Animal Science, Veterinary School, Universitat Aut\`onoma
de Barcelona, Bellaterra, 08193 Spain}
\affiliation{Key Laboratory for Animal Biotechnology of Jiangxi Province and the Ministry of
Agriculture of China, Jiangxi Agricultural University, Nanchang, 330045, China}

\author{Giacomo Marmorini}
\affiliation{Department of Physics, Tokyo University of Science, Tokyo 162-8601, Japan}
\affiliation{Research and Education Center for Natural Sciences,
Keio University,  Kanagawa 223-8521, Japan}
 
\author{Ginestra Bianconi}
\affiliation{ Department of Physics, Northeastern University, Boston 02115 MA, USA}

\begin{abstract}
Many complex networks from the World-Wide-Web to biological networks grow taking into account the heterogeneous features of the nodes. The feature of a node  might be a discrete quantity such as  a classification of a URL document as personal page, thematic website, news, blog, search engine,   social network, ect. or the classification of a gene in a functional module. Moreover the feature of a node  can be a continuous variable such as the position of a node in the embedding space. In order to account for these properties, in  this paper we provide a  generalization of growing network models with preferential attachment that includes the effect of heterogeneous features of the nodes. The main effect of heterogeneity is the emergence of an ``effective fitness'' for each class of nodes, determining the rate at which nodes acquire new links. The degree distribution exhibits a multiscaling behaviour analogous to the the fitness model. This property is robust with respect to variations in the model, as long as links are assigned through effective preferential attachment. 
Beyond the degree distribution, in this paper we give a full characterization of the other relevant properties of the model. We evaluate the clustering coefficient and show that it  disappears for large network size, a property shared with the Barab\'asi-Albert  model. Negative degree correlations are also present in this class of  models, along with non-trivial mixing patterns among features. We therefore conclude that both  small clustering coefficients and disassortative mixing  are outcomes of the preferential attachment mechanism in general growing networks.

\end{abstract}


\maketitle

\section{Introduction}

In the last ten years statistical mechanics has made great advances \cite{albert2002statistical,Caldarelli:2007,Latora_review,Barrat:2008,Newman:2010,Dorogovtsev:2009} in the understanding of the dynamics and the characteristic  structural properties  of complex networks. These findings shed light on the universal organization principles beyond  a large variety  of biological, social, communication and technological systems.
Recently, large attention as been  addressed to spatial networks  \cite{barthelemy2010spatial} in which the links are determined by the ``similarity'' or proximity of the nodes in the physical or hidden space in which the networks are embedded. 
Spatial networks \cite{barthelemy2010spatial} are found in communication \cite{krioukov2008efficient}, transportation\cite{bianconi2009assessing,Havlin} and even social networks \cite{Boguna}.
The role of space in complex networks significantly affects the dynamical properties of the graphs changing their navigability properties \cite{Kleinberg,krioukov2008efficient}, the critical behavior of the Ising model \cite{Bianconi_Ising} or the epidemic spreading \cite{barthelemy2010spatial}.

However, the similarity between two nodes might  also  correspond to  a modular organization  of the network \cite{Santo}. The general networks models that consider a modular structure are block-models \cite{Airoldi} and multifractal models \cite{Vicsek} that are intrinsically static models of modular networks. Nevertheless several networks are simultaneously growing and developing a modular structure.
For example, the World Wide Web contains pages falling into different classes with widely different features (personal pages, thematic websites, news, blogs, search engines, social networks\ldots) and links between any two pages are influenced by their qualities as well as by the specific classes to which they belong. 

Moreover many molecular networks, as for example protein-protein interaction networks, transcription networks and coexpression networks are scale-free \cite{Barabasioltvai}  and growing but have in addition a relevant  modular structure. In coexpression networks there is  a good correspondence between network modules and biological functions as characterized by Gene Ontology or pathway enrichment analyses \cite{stuart2003gene}. Connectivity between and within modules depends on tissue and species \cite{yang2011association} and should be taken into account in realistic models of coexpression networks. 
 
In light of these results it is necessary to understand how similarity,  spatial embedding, modular  structure or other features might change the nowadays classic description of growing networks following preferential attachment \cite{Barabasi1999emergence}. In this mechanism, new nodes are added at constant rate and connected to existing nodes of the network. The probability $\Pi(i)$ that a new node connects to a node $i$ 
is proportional to the degree $k_i$ of the node $i$, i.e. $\Pi(i)\propto k_i$. Preferential attachment mechanism has been directly measured for many complex  networks and remains a successful explanation for their scale-free degree distribution. Nevertheless, pure preferential attachment, as implemented in the Barab\'asi-Albert (BA) model \cite{Barabasi1999emergence}, has some drawbacks: for example, it generates a clustering coefficient that is small compared with the ones observed in real networks and follows a "first-mover-advantage" mechanisms to the extent that older nodes are systematically associated with larger degree. Several rules for network growth giving rise to an effective preferential attachment mechanism have been studied to overcome these limitations (see reviews in \cite{albert2002statistical,newman,krapivsky2000connectivity}).

Shortly after the seminal paper by Barab\'asi and Albert \cite{Barabasi1999emergence}, it was recognized that heterogeneity between nodes is an important ingredient for more realistic models, destroying the age-degree correlation present in the BA model. The first proposal in this direction was the addition of node quality or ``fitness'' to the BA model \cite{bianconi2001competition,bianconi2001bose}. This model by Bianconi and Barab\'asi paved the way for  the study of a wider class of models with other node features such as position in space  \cite{manna2002modulated,xulvi2002evolving,barthelemy2003crossover,santiago2007emergence,santiago2008extended}. Some properties of preferential attachment networks on metric spaces, such as the degree and link length distributions, were derived analytically by two of the authors \cite{ferretti2011}. The result is that both fitness \cite{bianconi2001competition} and space \cite{ferretti2011} give rise to networks with multi-scaling in the degree distribution (i.e., a sum of power laws).

In this paper we provide a general analysis of growing networks with both preferential attachment and \hvs. The preferential attachment probability from a new node to node $i$ is proportional to the degree $k_i$ of the node multiplied by a generic positive function of the \hvs\ $h$ of the new node and $h_i$ of the node $i$:
\beq
\Pi(i)\propto \pr(h_i,h)k_i
\eeq
The \hvs\ can have different interpretations: they can represent spatial embedding of the network, or discrete features of the nodes defining a modular structure of the network, or they can represent fitness describing the higher ability of some nodes to acquire new links. The space of feature and the connection function $\pr(h_i,h)$ are fixed and do not coevolve with the network.

We give a full account of the model determining the  degree distribution, clustering and assortativity. We derive the general expression for the asymptotic degree distribution in the rate equation approach. The form of this distribution is a convolution of power-laws depending on an "effective fitness" of the node, which is determined by the similarity matrix through a self-consistent equation.
We also discuss the conditions under which the rate equation approach breaks down and the fate of the network in these cases. 

We derive the general expression for the asymptotic clustering coefficient, showing that clustering always decreases as an inverse power of the network size and therefore disappears in the thermodynamic limit. Assortativity in node degree and \hvs\ are also studied in these networks, both numerically and analytically. Node degree correlations are negative as in the BA model: disassortativity  increases with the heterogeneity of the nodes and decreases slowly with the network size. 

Finally we allow for several variations  on the model, including addition and rewiring of links, and show that the results are robust with respect to these variations as long as the connections are assigned through preferential attachment.



\section{Degree distribution of growing networks with \hvs}
\label{costruzione}
\subsection{The model}\label{sectionmodel}
We present a general class of models for networks with preferential attachment and \hvs. In these models, each node has a \hv\ $h$ randomly chosen from a set $\Spa$ with probability $\pro(h)$. We call $h_i$ and $k_i$ the \hv\ and degree of the $i$th node. At each time step, a node with $m$ links is added to the network. These links are connected to existing nodes with probability 
\beq
\Pi(i)=\frac{\pr(h_i,h)k_i}{\sum_{j}
\pr(h_j,h)k_j}\label{prefattachrule}
\eeq
where $h$ corresponds to the \hv\ of the new node and $\pr(h',h)$ is a (positive) connection function from $h$ to $h'$. Such a model is therefore completely defined by the functions $\pro(h)$ and $\pr(h',h)$. 

In modular or community models, denoting by $N_c$ the number of communities,  $h$ is an integer number in $1\ldots N_c$ while $\pro(1)\ldots\pro(N_c)$ denote the relative sizes of the different communities. In the fitness model, $h$ is the quality of the nodes, that is, a real positive number. In spatial models, $h$ is the spatial position of the node; for example, in models on a plane, $h=(x,y)$ and $\pro(x,y)$ is the node density. 

\subsection{Degree distribution}
We derive the degree distribution through the rate equation approach introduced by Krapivsky, Redner and Leyvraz \cite{krapivsky2000connectivity}, modified as in \cite{ferretti2011}. The equation for $N_k(h)$, which is the average number of nodes with \hv\ $h$ and degree $k$, is
\begin{align}
N_k(h,t+1)&=N_k(h,t)+\delta_{k,m}\pro(h)-m\cdot \\ 
\cdot & \sum_{l\in\Spa}  \frac{\pr(h,l)(k N_k(h,t)-(k-1)N_{k-1}(h,t))}{\sum_{j\in\Spa} \pr(j,l)\sum_{k'=m}^\infty{k'} N_{k'}(j,t)} \pro(l)  \label{neq} \nonumber 
\end{align}
where the $\delta_{k,m}$ term in the left hand side  accounts for the birth of new nodes. If the \hvs\ are continuous, the number of nodes is substituted with the number density in the equation above, and the sums over the \hvs\ with integrals. 
More generally, if $\Spa$ 
is not a finite set, $p(h)$ denotes a probability measure over $\Spa$ and the sum $\sum_{h\in\Spa}p(h)f(h)$ should be read as $\int_S p(h)f(h)$, while $N_k(h,t)$ for fixed $t$ is a finite measure over $\Spa$.


We assume a linear scaling with time for the quantity
\beq
{\sum_{l\in\Spa} \pr(l,h)\sum_{k'=m}^\infty{k'}N_{k'}(l,t) }=mC(h)t+o(t)\label{linearscaling}
\eeq
where we neglect finite-size corrections contained in the $o(t)$ term, which depend on the initial nodes of the network \cite{cuenda2011simple}. Then we can define $n_k(h,t)=N_k(h,t)/t$ and rewrite it as 
\begin{align}
n_k(h,t+1)&\left(1+\frac{1}{t}\right)=n_k(h,t)+\frac{\delta_{k,m}\pro(h)}{t}+\label{eqn}\\ -&\frac{q(h)}{t}(k n_k(h,t)-(k-1)n_{k-1}(h,t))\nonumber
\end{align}
where $q(h)$ plays the same role as the (average) fitness of the node \cite{bianconi2001competition,bianconi2001bose} and is defined as 
\beq
q(h)=\sum_{l\in\Spa}  \frac{\pr(h,l) }{C(l)}\pro(l)\label{qdef}
\eeq
where $C(h)$ is determined by solving asymptotically the above equation (\ref{eqn}), obtaining
\begin{align}
{n_k(h)}&=\frac{\pro(h)}{q(h)}\frac{\Gamma(m+q(h)^{-1})\Gamma(k)}{\Gamma(k+1+q(h)^{-1})\Gamma(m)}\nonumber \\
&\simeq \frac{\pro(h)}{q(h)m}\left(\frac{k}{m}\right)^{-(1+q(h)^{-1})}\label{nkeq}
\end{align}
and substituting in the definition of $C(h)$ to obtain
\beq
C(h)=\sum_{l\in\Spa} \pr(l,h)\frac{\pro(l)}{1-q(l)}\label{ceq} 
\eeq
We can join (\ref{qdef}) and (\ref{ceq}) in a single functional equation for $q(h)$:
\beq
q(h)=\sum_{l\in\Spa}  \frac{\pr(h,l) \pro(l)}{\sum_{j\in\Spa} \pr(j,l)\frac{\pro(j)}{1-q(j)}}\label{qeq}
\eeq

From the point of view of the degree distribution, this class of models is equivalent to the fitness model of Bianconi and Barab\'asi \cite{bianconi2001competition,bianconi2001bose}, but in this case the fitness distribution is determined by $\pro(h)$ and $\pr(h,l)$ through equation (\ref{qeq}). The resulting degree distribution is
\beq
n_k=\sum_{h\in\Spa} n_k(h)\simeq\sum_{h\in\Spa}\frac{\pro(h)}{q(h)m}\left(\frac{k}{m}\right)^{-(1+q(h)^{-1})} \label{distrk}
\eeq 
so the distribution is a sum of power laws similarly to the fitness model, and for a regular distribution of $h$ and $q(h)$ it typically reduces to a power law with logarithmic corrections \cite{bianconi2001competition}. 

Actually, the fitness model itself is a particular example of such a model with a \hv\ $h=\eta\in [0,1]$ distributed as $\rho(\eta)$ and a connection probability $\pr(h,l)=h=\eta$. Then the  equation (\ref{qeq}) reduces to 
\beq
q(\eta)= \int_0^1 dy\  \frac{\eta }{\int_0^1 dz\ \frac{\rho(z)}{1-q(z)}z}\rho(y)= \frac{\eta }{\int_0^1 dz\ \frac{\rho(z)}{1-q(z)}z}\equiv \frac{\eta}{C}
\eeq
where $C$ depends on the whole distribution of $\eta$ and is determined by the usual consistency equation
\beq
1= {\int_0^1 dz\ \frac{\rho(z)}{C/z-1}}
\eeq
therefore the model reduces to the fitness model with $q(\eta)=\eta/C$. Also the spatial network models discussed in \cite{manna2002modulated,xulvi2002evolving,yook2002modeling,barthelemy2003crossover,santiago2007emergence,ferretti2011} are special cases of the above model, with the position playing the role of \hv.

Note that  since the sum of all node degrees $\sum_ik_i(t)$ should be equal to twice the total number of links $mt$, and since its mean is given by  $\langle\sum_ik_i(t)\rangle=\sum_hmp(h)t/(1-q(h))$, 
$q(h)$ should satisfy an additional identity
\beq
\sum_{h\in\Spa} \frac{p(h)}{1-q(h)}=2\label{qid2}
\eeq
However, this identity is not new but can be derived from (\ref{qdef}) and (\ref{ceq}) by substituting the definition of $C(h)$ in the numerator of the identity $1=\sum_hp(h)C(h)/C(h)$ and rearranging.  

Interestingly, many simple models are based on ``symmetric'' \hvs, that is, all the values of the \hvs\ are equivalent. In other terms, there is a group of bijective transformations $T_\alpha$ from $\Spa$ to itself such that the distribution and connection function are invariant (that is, $\pro(T_\alpha(h))=\pro(h)$ and $\pr(T_\alpha(h),T_\alpha(h'))=\pr(h,h')$) and moreover the action of the group is transitive (that is, for every pair $h,h'$ there is a transformation $T_\beta$ mapping $h$ in $h'$, $T_\beta(h)=h'$). We denote these models as homogeneous models. In this case, if all the sums in the above equations are convergent, the symmetry implies $q(T(h))=q(h)$ and transitivity implies that all $q$ are the same, then from equation (\ref{qid2})  we obtain immediately $q=1/2$. This means that all homogeneous models have the same degree distribution $n_k\sim k^{-3}$ of the Barab\'asi-Albert model. We will see some examples in section \ref{examples}. 
We can actually extend the argument to a slightly more general condition, following \cite{jordan2010}: if the quantities $\sum_{l\in\Spa} p(l)\pr(l,h)$ and $\sum_{l\in\Spa} p(l)\pr(h,l)$ are equal and independent of $h$, then the degree distribution is the same of the BA model.

For non-homogeneous models, the equations (\ref{qeq}) often need to be solved numerically. For discrete features, a solution can be obtained by root-finding methods. For continuous features, an effective way to solve this kind of equations  was presented in \cite{ferretti2011}.


\subsection{Examples}\label{examples}


\subsubsection{Bipartite networks}

A bipartite network is usually composed by two classes of nodes ($h=1,2$) with links connecting only nodes of different classes. To build a scale-free bipartite network, we choose $\sigma_{(1,1)}=\sigma_{(2,2)}=0$. We assume that a node can belong to class $1$ or $2$ with probabilities $p_1$, $p_2$ (with $p_1+p_2=1$). Note that in this model the non-zero terms of the connection function can be redefined as $\sigma_{(1,2)}=\sigma_{(2,1)}=1$, so the function is actually symmetric. 

The qualities of the two classes from equation (\ref{qeq}) are then 
\beq
q_1=p_2\quad,\quad q_2=p_1
\eeq and the degree distribution is given by $n_k=\frac{p_1}{m(1-p_1)}\left(\frac{k}{m}\right)^{-(2-p_1)/(1-p_1)}+\frac{(1-p_1)}{mp_1}\left(\frac{k}{m}\right)^{-(1+p_1)/p_1}$. Note that if $p_1=p_2=1/2$, the model is actually homogeneous and $n_k\sim k^{-3}$ as expected from our general arguments.

Similar models have been applied to the human sexual networks, which have a bipartite heterosexual component \cite{ergun2002human}.

\subsubsection{A network with asymmetric connection function}

Another simple but interesting example can be obtained by assuming two kinds of nodes, ``central'' and ``periferic''  (with probabilities $p_C$ and $p_P=1-p_C$) such that two periferic nodes are never connected (i.e. the 
connection function satisfies $\sigma_{(P,P)}=0$). The other terms of the connection function can be always redefined as $\sigma_{(C,C)}=\sigma_{(C,P)}=1$ so its only parameter is $\bar{\sigma}\equiv\sigma_{(P,C)}\in[0,+\infty]$. This parameter controls the asymmetry of the connection function.

The relevant solution of equation (\ref{qeq}) with this connection function is 
\begin{align}
q_C &= 1-\frac{2(1-\bar{\sigma})p_C}{1+p_C-3\bar{\sigma}+\sqrt{(1+p_C-3\bar{\sigma})^2+8\bar{\sigma}(1-\bar{\sigma})}} \\
q_P & =\bar{\sigma}\cdot\frac{p_C-1-\bar{\sigma}+\sqrt{(1+p_C-3\bar{\sigma})^2+8\bar{\sigma}(1-\bar{\sigma})}}{1+p_C-3\bar{\sigma}+\sqrt{(1+p_C-3\bar{\sigma})^2+8\bar{\sigma}(1-\bar{\sigma})}}  
\end{align}
In the limit $\bar{\sigma}\rightarrow\infty$, the model reduces to a bipartite network and the solution to $q_C=1-p_C$, $q_P=p_C$ as expected.


\subsubsection{Community structure}
\label{commstruct}
To model scale-free networks with community structure \cite{girvan2002community}, a simple possibility is to label each community by a \hv\ $h_c$ and choose a connection function with $\sigma(h_c,h_c)>\sigma(h_c,h_{c'})$. In the simplest model, all 
communities have the same size and connect randomly to the other communities, with some preference for self-connections. (The corresponding connection function is $\sigma(h_c,h_{c})=1$ for all $h_c$, while $\sigma(h_c,h_{c'})=\bar{\sigma}$ for all pairs $h_c\neq h_{c'}$.) In this case the model is actually homogeneous and has the same degree distribution as the BA model, independently on the number of communities and the value of $\bar{\sigma}$.

\subsubsection{Hierarchical structure and navigable networks}

Navigable networks are often based on a hierarchical structure \cite{newmanwatts}, with a connection probability that depends on the distance on a tree representing the hierarchical levels and the nodes. Scale-free navigable networks can be easily built by choosing a symmetric connection function depending only on the distances on the tree. (Note that these models are actually spatial models, since a tree is an ultrametric space.)

Even if there is a lot of interesting structure in these networks, the degree distribution follows the simple multi-scaling behaviour in equation (\ref{distrk}). In particular, the simplest cases of binary or n-ary trees (or more general trees where the length and the number of branches splitting from a single branch depend only on the level), with nodes located at the top of the terminal branches, have the usual degree distribution $n_k\sim k^{-3}$, since these trees are homogeneous spaces. 

\subsubsection{Modular structure with fitness}

As a final example, we discuss a model with both modular structure and fitness. This model can be considered a simplified model of the WWW. We assume that each page is assigned to some category $n$ according to type, content and functionality. Each category could have a different relative size $\pi_n$ and a distribution of page fitness $\rho_n(\eta)$. Moreover, the relative importance of different categories for a page of category $q$ is given by the weights $w_{n,q}$, with $\sum_q w_{n,q}=1$. (For a page of category $n$, the weights $w_{n,q}$ affect its probability of being linked by pages from other categories.) The network evolves as follows: at each time, a new node is added to the network and assigned to the $n$th category with probability $\pi_n$, then its fitness is randomly extracted from $\rho_n(\eta)$. The node is connected to the existing nodes according to a probability proportional to the fitness $\eta$, the weight $w$ and the degree $k$ of the nodes:
\beq 
\Pi(i)=\frac{ \eta_i w_{n_i,n} k_i}{\sum_j \eta_j w_{n_j,n} k_j}
\eeq
so the model dynamics follows equation (\ref{prefattachrule}) with $h=(n, \eta)$, $\sigma(h_i,h)=\eta_i w_{n_i,n}$ and $p(h)=\pi_n \rho_n(\eta)$. Similar models appear in a natural way in the study of many systems. In this model, beyond the fitness, the dynamics of a node is influenced by the relative size of its category $\pi_n$ as well as by the weights $w_{n,q}$ and the sizes of other categories $\pi_q$. In the WWW example, there are millions of blogs but only a few search engines, and there is a good probability that a blog links a search page; this explains the different connectivity and degree distribution of these categories.

\begin{figure}
\includegraphics[width=0.45\textwidth]{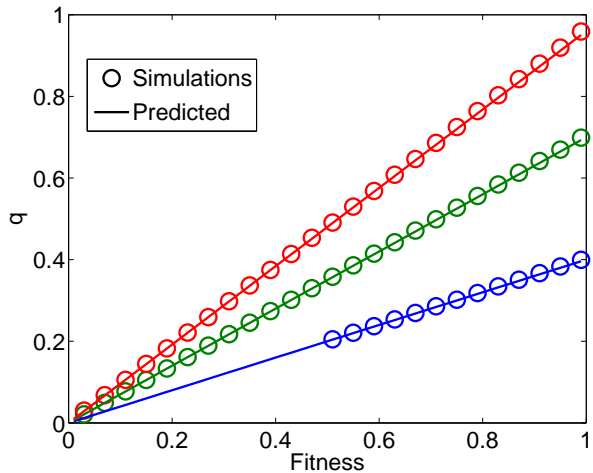}
\caption{(Color online) Node qualities as a function of fitness in a model with hierarchical modules A,B,C and fitness. Continuous lines represent the theoretical predictions. The hierarchy is A$<$B$<$C and nodes in each module can connect only to nodes in the same or higher modules with equal probabilities. The fitness distributions are $\rho_A(\eta)=U_{(1/2,1)}(\eta)$, $\rho_B(\eta)=U_{(0,1)}(\eta)$ and $\rho_C(\eta)=2\eta$. Simulated networks have size $N=10^7$ and initial node degree $m=5$. Qualities are measured as $\hat{q}=\log\left(\left\langle k(t)\right\rangle/\left\langle k(t/2)\right\rangle\right)/\log 2$ as detailed in \cite{ferretti2011}. }
\label{fig_modfitness} 
\end{figure}

The node qualities for this model from equation (\ref{qeq}) are equivalent to a modified fitness model 
$q(\eta,n)={\eta}/{\gamma_n}$, where the coefficients $\gamma_n$ solve the nonlinear equations
\beq
\frac{1}{\gamma_n}=\sum_m w_{n,m}\pi_m\left(\sum_l w_{l,m}\pi_l\int d\eta \frac{\rho_l(\eta)}{\eta^{-1}-\gamma_l^{-1}}\right)^{-1}
\eeq
Simulation results are in very good agreement with the numerical solution of these equations, as shown in Figure \ref{fig_modfitness}.

\subsection{Breakdown of the rate equation approach}

If the space $\Spa$ of \hvs\ has a finite number of elements, the approach presented here gives rise to a finite number of nonlinear equations (\ref{qdef}),(\ref{ceq}) in the variables $q(h),C(h)$. If these equations admit a single solution, as we expect, then the degree distribution follows equation (\ref{distrk}).  

If the number of \hvs\ is infinite, equations (\ref{qdef}),(\ref{ceq}) could involve divergent sums (or integrals). In particular, there are two situation where the rate equations break down:
(i) the sums of the connection function are divergent, that is, the connection function is not measurable, or (ii) all sums converge but there is no solution to the selfconsistency equations. We discuss these two scenarios in the next sections.   

\subsubsection{Heterogeneity
-driven attachment}
If the sums of the connection function are divergent, links attach preferentially to some nodes not chosen on the basis of preferential attachment but belonging to the sets of \hvs\ with divergent sums of the connection function. This can give rise to exponential tails or condensation or other behaviour, depending on form of the connection function. As an example, spatial models with a divergent connection function near $d=0$ show a behaviour similar to nearest-neighbour attachment and consequently an exponential tail \cite{manna2002modulated,ferretti2011}. Another example is given by networks on a flat space of dimension $D$ with uniform node density and a connection function $\sigma(x,x')$ that is an inverse power law in the distance between the node and a point $\hat{x}$, for example $\sigma(x,x')= d(x,\hat{x})^{-k}$ with $k>D$. In this case the divergence is localized around $\hat{x}$, prompting condensation on the nodes closer to $\hat{x}$ since these nodes get most of the new connections.   

\subsubsection{Bose-Einstein condensation}
Even if all the sums of the connection function are convergent, it is still possible to find cases where the equations for $q(h)$ do not admit solutions. In fitness models, the lack of solution of the self-consistent equations is a signal of Bose-Einstein condensation of links \cite{bianconi2001bose} on the nodes of highest fitness. 

In more general models, condensation occurs on nodes close to a \hv\ $h_c$ determined as follows. Each $h\in\Spa$ corresponds to an element of the set $M$ of measures $\mu_h(h')=\pr(h,h')p(h')$. Conversely, given a Radon metric on $M$, each point in its closure $\bar{M}$ can be mapped to a \hv\ belonging to an ``extended'' space $\bar{\Spa}$. The Radon metric on $\bar{M}$ induces a metric on $\bar{\Spa}$, which can therefore be thought of as the closure of $\Spa$: the elements of $\bar{\Spa}$ are ``limit points'' of $\Spa$ and are the candidates for condensation. In particular, the location $h_c$ of the condensate can be found by considering the addition of a single node with variable $\bar{h}\in \bar{\Spa}$ to the network and maximizing its asymptotic link share $n_{\bar{h}}$:
\beq
h_c=\argmax_{\bar{h}\in \bar{\Spa}} n_{\bar{h}}
\eeq

As a simple example, in the fitness model with $\Spa=[0,\eta_{max})$, the set $M$ contains the measures $\mu_\eta(\eta')=\eta p(\eta')$ and its closure corresponds to $\bar{\Spa}=[0,\eta_{max}]$, which is the closure of $\Spa$. In this case $n_{\bar{\eta}}$ is zero unless $\bar{\eta}=\eta_{max}$, because the consistency equation with an additional node has no finite solution for other values of $\bar{\eta}$, therefore condensation occurs near $\eta_{max}$ as expected.


 

\section{Clustering and assortativity}

\subsection{Clustering}
We define the average clustering coefficient $C_{clust}$ of a network as $C_{clust}=3\cdot n_\mathrm{triangles}/ n_\mathrm{triples}$. It is well known that the clustering coefficient of the Barab\'asi-Albert model decreases as the network size $\N$ increases, converging to zero in the thermodynamic limit \cite{bollobas}. As we show in the next sections, this property is quite general, being shared by all heterogenous models under some conditions on the convergence of sums of $\pr(h,h')$. This mean that heterogeneity or features cannot account for non-vanishing clustering coefficients observed in most real networks.

\subsubsection{Clustering in the Barab\'asi-Albert and in homogeneous model}
Both the average number of triangles and the average number of triples can be easily computed from the preferential attachment rule if we assume that node degrees follow the continuum equation for the mean degree $k_i(t)=m(t/t_i)^{q(h_i)}$ \cite{Barabasi1999mean}. This approach has been applied to the BA model in \cite{fronczak2003mean}. Here we generalize it to models with features. First, we summarize the computation for the BA case. The asymptotic number of triangles is given by
\beq
n_\mathrm{triangles}\sim\frac{m^2(m-1)\ln^3\N}{48}
\eeq
Denoting by $t_A<t_B<t_C$ the birth times of a triplet of nodes, BA networks contain three kind of triples: $A\leftarrow B\leftarrow C$, $A\leftarrow C\rightarrow B$ and $B\rightarrow A\leftarrow C$. Since older nodes have the highest degrees and are therefore the most attractive under preferential attachment, for large $\N$ almost all triples are of the last kind and their number is given by
\beq
n_\mathrm{triples}\sim\frac{m^2\N\ln \N}{2}
\eeq
therefore obtaining the known result for the asymptotic clustering coefficient 
\beq
C^{BA}_{clust}\sim \frac{(m-1)}{8}\frac{\ln^2 \N}{\N}\label{eqclustba}
\eeq
Details of the calculations 
can be found in appendix \ref{clustba}.

For homogeneous models ($q(h)=1/2$) the same calculation is valid for the number of triples, while the number of triangles and therefore the clustering coefficient are multiplied by a factor dependent on the connection function:
\beq
\frac{C_{clust}}{C_{clust}^{BA}}= \left\langle \frac{\pr(h_A,h_B)\pr(h_A,h_C)\pr(h_B,h_C)}{C(h_B)C(h_C)^2/8}\right\rangle_{A,B,C}\label{eqclusthom}
\eeq
Features 
appear only in this factor. For example, bipartite networks have no triangles and the above factor is zero, giving $C_{clust}=0$. Note that the above factor could be divergent: in this case our approach breaks down and  the asymptotic behaviour of $C_{clust}(\N)$ could change. However, we are not aware of any model of this kind  with non-zero clustering in the limit $\N\rightarrow \infty$ and a power-law tail in the degree distribution. 

For a symmetric model with community structure, the clustering coefficient is
\beq
\frac{C_{clust}^{com}}{C_{clust}^{BA}}= 
\frac{N_c+3N_c(N_c-1)\bar{\sigma}^2+N_c(N_c-1)(N_c-2)\bar{\sigma}^3}{(1+\bar{\sigma}(N_c-1))^3}
\eeq

\begin{figure}
\includegraphics[width=0.450\textwidth]{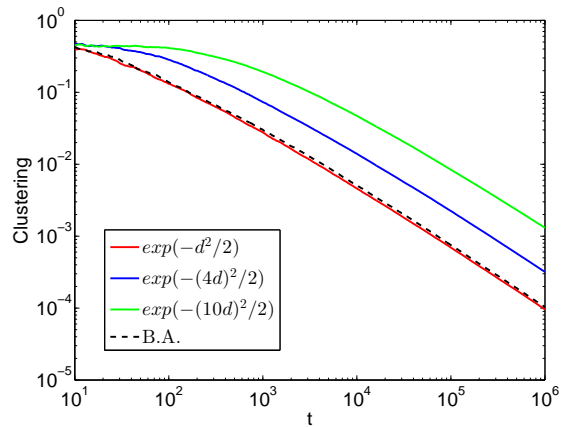}
\caption{(Color online) Plot of the clustering coefficient as a function of the size of the network, for networks on the two-dimensional flat disk of radius $r=1$ with Gaussian connection functions.}
\label{fig_clust} 
\end{figure}

An additional remark applies to spatial networks with short-range interactions and networks with community structure. At short times, these models resemble geometric random graph models,  which exhibit strong clustering \cite{barthelemy2003crossover}. However, when the average number of nodes in the interaction volume increases, the clustering coefficient decreases as a consequence of the preferential attachment dynamics. At longer times, clustering falls according to eqs. (\ref{eqclustba}),(\ref{eqclusthom}) but remains higher for spatial networks than for the Barab\'asi-Albert model, as it is shown in Figure \ref{fig_clust}.

\subsubsection{Clustering in general models}

For general models with some $q(h)\neq 1/2$, the calculation is similar to the homogeneous case. The dominant contribution comes from triples and triangles with a vertex of maximum fitness $q_M=q(h_M)>1/2$  with $h_M=\argmax_{h\in \Spa}{q(h)}$. The leading contribution to the average number of triangles is
\begin{align}
& n_\mathrm{triangles}\sim \frac{m^2(m-1)p(h_M)}{(2q_M-1)^2} \N^{2q_M-1}\ln \N\cdot \\ & \cdot \sum_{h_B,h_C}\frac{\pr(h_M,h_B)\pr(h_M,h_C)\pr(h_B,h_C)p(h_B)p(h_C)}{C(h_B)C(h_C)^2}\nonumber
\end{align}
while the number of triples is
\beq
n_\mathrm{triples}\sim\frac{m^2p(h_M)}{2(2q_M-1)}\N^{2q_M}
\eeq
then the clustering coefficient is
\begin{align}
C_{clust}&\sim \frac{6(m-1)}{2q_M-1}\frac{\ln \N}{\N}\cdot \\ \cdot&\sum_{h_B,h_C}\frac{\pr(h_M,h_B)\pr(h_M,h_C)\pr(h_B,h_C)p(h_B)p(h_C)}{C(h_B)C(h_C)^2} \nonumber
\end{align}
This expression is valid as long as the sum inside it is finite. So the asymptotic clustering is sligthly smaller than in the BA model. 

From this general result we can extract the clustering for the fitness model of Bianconi and Barab\'asi \cite{bianconi2001competition} in the fit-get-rich phase:
\beq
C_{clust}^{BB}\sim \frac{6(m-1)\langle \eta\rangle}{C^2(2-C)}\frac{\ln \N}{\N}
\eeq
assuming $\eta_{max}=1$ and $\rho(\eta_{max})>0$.

\subsection{Assortativity: \hvs}
Nodes with different $h$ are not randomly connected: in- and out-going links connect preferentially nodes with some features. These preferences are embedded in the in- and out-distributions $f^{IN}(h_i,h')$ and $f^{OUT}(h_i,h')$, which are asymptotically defined by the equations 
\beq
k^{IN}_{i(h')}=f^{IN}(h_i,h')k^{IN}_i \label{fin} 
\eeq
\beq
k^{OUT}_{i(h')}=f^{OUT}(h_i,h')k^{OUT}_i \label{fout} 
\eeq
where $k^{IN}_i$ and $k^{OUT}_i$ are the number of in- and out-going links for the $i$th node (note that in these models $k^{OUT}_i=m$), while $k^{IN}_{i(h')}$ and $k^{OUT}_{i(h')}$ are the numbers of in- and out-going links between  the $i$th node and nodes with variable $h'$. From the definitions above, we have $\sum_{h'}f^{IN}(h_i,h')=\sum_{h'}f^{OUT}(h_i,h')=1$.

The distributions $f^{IN}$ and $f^{OUT}$ are positive. $f^{IN}$ can be obtained from the continuum equations for $k^{IN}_{i(h')}$
\beq
\frac{dk^{IN}_{i(h')}}{dt}=\frac{\pr(h_i,h')p(h')(k^{IN}_i+m)}{C(h')t}
\eeq
by plugging in equation (\ref{fin}) and comparing with the continuum equation for $k^{IN}_{i}$
\beq
\frac{dk^{IN}_{i}}{dt}=q(h_i)\frac{k^{IN}_i+m}{t}
\eeq
giving as a result  
\beq
f^{IN}(h_i,h')=\frac{\pr(h_i,h')p(h')}{C(h')q(h_i)}
\eeq
while $f^{OUT}$ can be obtained from equation (\ref{prefattachrule}) by substituting $k_i$ with the mean of the total degree of nodes of feature $h'$:
\beq
f^{OUT}(h_i,h')=\frac{\pr(h',h_i)p(h')}{C(h_i)(1-q(h'))}
\eeq

From these distributions it is easy to find the fraction of links between nodes with variables $h', h''$:
\begin{align}
\varphi(h',h'') & =  \frac{p(h')p(h'')}{1+\delta_{h',h''}}\cdot \label{assort} \\
\cdot & \left(\frac{\pr(h',h'')}{C(h'')(1-q(h'))}+  \frac{\pr(h'',h')}{C(h')(1-q(h''))} 
\right) \nonumber 
\end{align}
that should be compared with the null value of the same quantity, obtained by a random re-arrangement of links that preserves degree (i.e., unassortative connections):
\beq
\varphi_{\NULL}(h',h'')=\frac{p(h')p(h'')}{1+\delta_{h',h''}}\frac{1}{2(1-q(h'))(1-q(h''))}
\eeq

As an example, consider the simplest model with community structure in section \ref{commstruct}. Denote the number of communities by $N_c$ and the only parameter of the connection function by $\bar{\sigma}=\pr(h,h')/\pr(h,h)$ for $h'\neq h$. The null distribution of links is given by $\varphi_{\NULL}(h',h')=1/N_c^2$, $\varphi_{\NULL}(h',h'')=2/N_c^2$ for $h''\neq h'$, while the actual assortativity depends on the parameter $\bar{\sigma}$: 
\beq
\varphi(h',h'')=\begin{cases}\frac{1}{N_c^{2}}\cdot\left[ 1+(\bar{\sigma}-1)\frac{N_c-1}{N_c} \right]^{-1}\quad \mathrm{if}\ h'=h'' \\ 
\frac{2}{N_c^{2}}\cdot\left[1+\left(\frac{1}{\bar{\sigma}}-1\right)\frac{1}{N_c} \right]^{-1}\quad \mathrm{if}\ h'\neq h''
\end{cases}
\eeq
so links are randomly distributed between \hvs\ if $\bar{\sigma}=1$, while the mixing is assortative (that is, $\varphi(h,h)/\varphi_{\NULL}(h,h)>\varphi(h',h'')/\varphi_{\NULL}(h',h'')$) for $\bar{\sigma}<1$ and disassortative for $\bar{\sigma}>1$.  
  
For the Bianconi-Barab\'asi fitness model, the distributions are 
\begin{align}
\varphi_{\NULL}(\eta,\eta')&=\frac{\rho(\eta)\rho(\eta')}{2\left(1-\frac{\eta}{C}\right)\left(1-\frac{\eta'}{C}\right)}\\
\frac{\varphi(\eta,\eta')}{\varphi_{\NULL}(\eta,\eta')}&= 2 \left[\frac{\eta}{C}\left(1-\frac{\eta'}{C}\right) + \frac{\eta'}{C}  \left(1-\frac{\eta}{C}\right) \right]        
\end{align}
and therefore the system shows disassortative mixing, since the ratio ${\varphi(\eta,\eta')}/{\varphi_{\NULL}(\eta,\eta')}$ is higher between high and low fitness than between similar fitnesses. 


\subsection{Assortativity: degree correlations}

Nontrivial connection functions $\pr(h,h')$ and node distributions $p(h)$ do not only induce assortative mixing between nodes with different variables, but affect also the degree-degree correlations between neighbours, as shown in \cite{krapivsky2001organization} for the BA model. The BA model is disassortative \cite{newman2003mixing}; however, in models with \hvs, the pattern of (disassortative) mixing between nodes with different degree is influenced by the hidden space and the connection function.

\subsubsection{Analytical results: continuum approximation}
A first understanding of degree correlations can be obtained by assuming deterministic evolution of node degree $k=m(t/t_0)^q$. In this approximation, average nearest-neighbour degree is given by
\beq
\langle k_{NN}(k)\rangle=\sum_h\frac{n_k(h)}{\sum_jn_k(j)}\sum_i\int_m^\infty dk_{NN}\ k_{NN} P(k_{NN},i|k,h) 
\eeq
where $P(k_{NN},i|k,h)$ is the probability that if we choose a random neighbour of a random node with features $h$ and degree $k$, the node chosen has degree $k_{NN}$ and features $i$. If we denote  the birth times of these nodes by $t_0$ and $t_{0NN}$ respectively, then if $t_{0NN}<t_0$  (i.e. $(k_{NN}/m)^{1/q(i)}>(k/m)^{1/q(h)}$) this probability can be obtained from the connection probability $m \frac{\sigma(i,h)k_{NN}(t_0)}{mC(h)t_0}$ multiplied by $1/k$ (which accounts for the random neighbour chosen), $p(i)$ (the probability that a node of feature $i$ is born at time $t_{0NN}$) and $dt_{0NN}/dk_{NN}$ (the Jacobian of the mapping from $k_{NN}$ to $t_{0NN}$). The final result is given by $\frac{dt_{0NN}}{dk_{NN}}\frac{p(i)\sigma(i,h)k_{NN}(t_0)}{kC(h)t_0}$. In the other case, if $t_{0NN}>t_0$, the relevant connection probability is $m \frac{\sigma(h,i)k(t_{0NN})}{mC(i)t_{0NN}}$. Substituting the present values of $k$ and $k_{NN}$, we obtain 
\beq
P(k_{NN},i|k,h)=\begin{cases}
\frac{\sigma(i,h)p(i)}{mq(i)C(h)}  (k/m)^{\frac{1-q(i)-q(h)}{q(h)}} (k_{NN}/m)^{-1/q(i)} \\
\qquad\qquad \mathrm{for}\ k_{NN}>m\left(\frac{k}{m}\right)^{q(i)/q(h)}\ ;\ \\ 
\frac{\sigma(h,i)p(i)}{mq(i)C(i)} (k_{NN}/m)^{-1-q(h)/q(i)} \\
\qquad\qquad \mathrm{for}\  k_{NN}<m\left(\frac{k}{m}\right)^{q(i)/q(h)}\ .\  
\label{asscont}       
\end{cases}
\eeq
For the BA model, we have $P(k_{NN}|k)=m k_{NN}^{-2}$ and the average nearest neighbour degree $\langle k_{NN}(k)\rangle=m\log(t)/2$ is independent of $k$, so the model shows no assortativity at all in this approximation. On the other hand, models with features  tend to be disassortative, i.e. $\langle k_{NN}(k)\rangle$ decreases with $k$. However, this is not true for all classes of nodes: for example, low quality nodes tend to be assortative, i.e. they attach more often to low quality nodes with similar degree.

\begin{figure}
\includegraphics[width=0.45\textwidth]{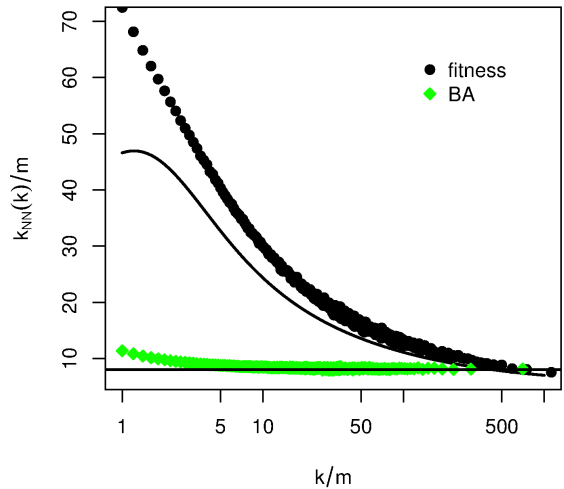}
\caption{(Color online) Plot of $\langle k_{NN}(k)\rangle$ for the BA model and for a fitness model with $\rho(\eta)=\frac{1}{2}\delta(\eta-1/2)+ \frac{1}{2}\delta(\eta-1)$.  Results are averaged from 20 simulations of size $N=10^7$ and initial degree $m=5$. Continuous lines correspond to the predictions from continuum equation (\ref{asscont}).}
\label{fig_ass} 
\end{figure}

In figure \ref{fig_ass} we compare the above results (\ref{asscont}) with the actual values of $\langle k_{NN}(k)\rangle$ in the BA model and in a simple fitness model.  The continuum approach describes correctly the qualitative pattern of degree correlations in these models, even if it does not fully account for their disassortativity.

It is also possible to estimate the asymptotic behaviour of the assortativity coefficient in the same approximation. We denote the maximum quality by $q_M=\max_h q(h)$. As in the BA model, the coefficient $\cass$ tends to zero in the thermodynamic limit $t\rightarrow\infty$. However, its asymptotic behaviour is $\cass\sim t^{q_M-1}$, and since $q_M\simeq 1$ in many models with features, its decrease is typically very slow compared to the BA model ($\cass\sim t^{-1/2}$).

\subsubsection{Analytical results: rate equation approach}
Assortativity in node degree can be easily estimated from the matrix $n_{k,l}$, defined as the fraction of links in the network joining a node of degree $k$ with a younger node of degree $l$. In models with \hvs, we define $n^{(h,i)}_{k,l}$ as the fraction of links joining a node of degree $k$ and variable $h$ with a younger node of degree $l$ and variable $i$. The matrix $n^{(h,i)}_{k,l}$ takes into account correlations between both degrees and \hvs.  
Asymptotically, this matrix can be obtained as in \cite{krapivsky2001organization}, by solving the difference equation 
\begin{align}\label{eqass}
n^{(h,i)}_{k,l} & (1+q(h)k+q(i)l)=q(h)(k-1)n^{(h,i)}_{k-1,l}+\\
+ & q(i)(l-1)n^{(h,i)}_{k,l-1}
+\delta_{l,m}\frac{\sigma(h,i)p(i)}{C(i)}(k-1)n_{k-1}(h)\nonumber 
\end{align}
This equation implies that $n^{(h,i)}_{k,l}\propto F(q(h),q(i))\cdot {\sigma(h,i)p(i)p(h)}/{C(i)}$ where $F(q(h),q(i))$ is a  function of the qualities, in agreement with the form (\ref{assort}) for the assortative mixing between \hvs. The matrix $n_{k,l}$ can then be obtained as $n_{k,l}=\sum_{h,i}n_{k,l}^{(h,i)}$. 

In principle, the equation (\ref{eqass}) can be solved by generating function methods discussed in appendix \ref{analyticassort}. The general solution for $m=1$ is
\begin{align}
n^{(h,i)}_{k,l} & = \frac{\sigma(h,i)p(i)}{C(i)q(h)} \sum_{c=2}^\infty (c-1) n_{c-1}(h)\cdot \nonumber \\  
\cdot & \sum_{j=0}^{l-1}(-1)^j \frac{1}{\alpha_j}{ l-1 \choose j} \sum_{n=0}^{k-c} (-1)^n \cdot \nonumber \\
\cdot & \frac{\Gamma(\alpha_j+1+k-c-n)}{(\alpha_j+k-n)\Gamma(\alpha_j-n)\Gamma(n+1)\Gamma(k-c-n+1)}\label{gensolass}
\end{align}
where $\alpha_j=\frac{1+(j+1)q(i)}{q(h)}$. However, extracting information from this solution is difficult.

The existence of degree correlations can be also shown simply by looking at the scaling of the quantity $n^{(h,i)}_{k,l}$ for $k\gg l$ and $k\ll l$. In the BA model with $m=1$, the scaling is  $n_{k,l}\sim k^{-2} l^{-2}$ and $k l^{-5}$ respectively \cite{krapivsky2001organization}, which is different from the naive $k^{-2}l^{-2}$ expected in the absence of degree correlations. 

The scaling $k^{-2} l^{-2}$ for $k\gg l$ can be understood from the following simple arguments. The degree of young nodes is dominated by outgoing connections, which select random nodes with probability proportional to $kn_k\sim k^{-2}$. On the other way, the attractiveness of old nodes decays with time as $t^{-1/2}$, therefore the distribution of the linked nodes (assuming a deterministic evolution $\bar{k}=l(t)\sim t^{-1/2}$ as a function of the birth time $t$) is $dt(l)/\sqrt{t(l)}=dl/l^2$.

To obtain the scaling for $n^{(h,i)}_{k,l}$ in these models with $m=1$, we approximate the difference equation with the corresponding differential equation $n^{(h,i)}_{k,l}=q(h)\partial (kn^{(h,i)}_{k,l})/\partial k+q(i)\partial (l n^{(h,i)}_{k,l})/\partial l$ and match the solution with the exact boundary conditions $n^{(h,i)}_{k,1}$ (for $k\gg l$) and $n^{(h,i)}_{2,l}$ (for $l \gg k$). The computation is outlined in appendix \ref{analyticassort}. The result 
is 
\beq
n^{(h,i)}_{k,l}\sim \begin{cases}
k^{-1/q(h)}l^{-1-q(h)/q(i)}\quad \mathrm{for}\ k\gg l \\ 
kl^{-1-(1+2q(h))/q(i)}\quad \mathrm{for}\ k\ll l \end{cases}
\eeq 
which is quite different from the null scaling $k^{-1/q(h)}l^{-1/q(i)}$. This shows that degree correlations exist also in heterogeneous model. Note that the scaling for $k\gg l$ is consistent with the continuum approximation (\ref{asscont}). The same scaling for $k\gg l$ and $l$ fixed can also be found directly from the generating function (\ref{gfass}) using Tauberian theorems. For $m>1$, the scaling for $k\gg l$ is the same as for $m=1$, while the scaling for $l\gg k$ changes to $n^{(h,i)}_{k,l}\sim k^ml^{-1-(1+(m+1)q(h))/q(i)}$. 

The BA model is slightly disassortative in degree and the addition of features does not change this property. This is already apparent from the numerical results in the previous sections, but can be also understood from the above scaling properties. In fact, in the BA model the asymptotic ratio between $n_{k,l}$ and its null value scales between $1$ for $k\gg l$ and $(k/l)^3\sim 0$ for $k\ll l$, therefore decreasing while $k$ and $l$ get closer. In homogeneous models ($q=1/2$), like the symmetric communities model, the scaling is the same as in the BA model. On the other side, in models with different qualities (like the Bianconi-Barabasi fitness model) the pattern of degree correlations is nontrivial, as already observed in the previous section: for example, old and well-connected nodes with features $h$ are preferentially linked to younger nodes of features $i$ and similar degree if qualities are low ($q(i)+q(h)<1$), but they link instead to younger nodes of low degree if qualities are high ($q(i)+q(h)>1$).



\subsubsection{Numerical results: spatial networks}

\begin{figure}
\includegraphics[width=0.450\textwidth]{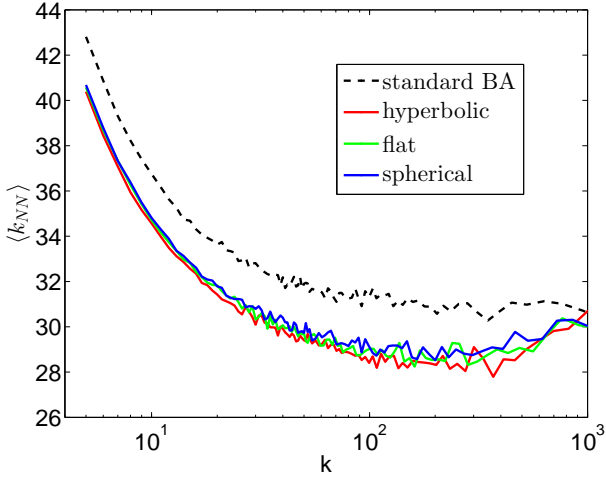}
\caption{(Color online) Plot of average neighbour degree $\langle k_{NN}(k)\rangle$ (averaged over 50 simulations) for networks of size $2\cdot 10^5$ on disks of radius $r=1.5$ in two-dimensional spaces of curvature $+1,0,-1$ with connection function $\tilde{\sigma}(d)=e^{-16d/r}$.}
\label{fig_exp16} 
\end{figure}

\begin{figure}
\includegraphics[width=0.45\textwidth]{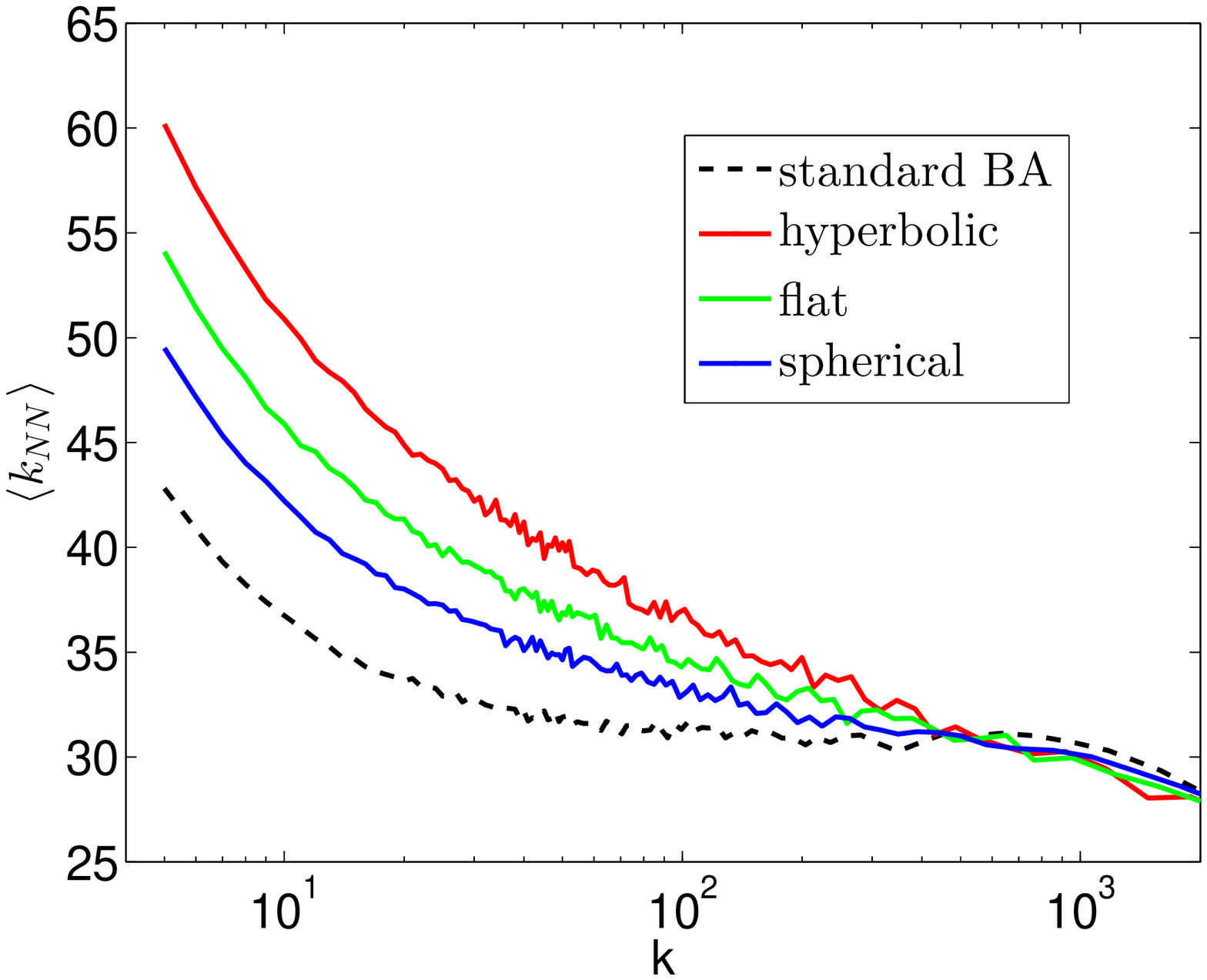}
\caption{(Color online) Plot of average neighbour degree $\langle k_{NN}(k)\rangle$ (averaged over 50 simulations) for networks of size $2\cdot 10^5$ on disks of radius $r=1.5$ in two-dimensional spaces of curvature $+1,0,-1$ with connection function $\tilde{\sigma}(d)=e^{-4d/r}$.}
\label{fig_exp4} 
\end{figure}

\begin{figure}
\includegraphics[width=0.45\textwidth]{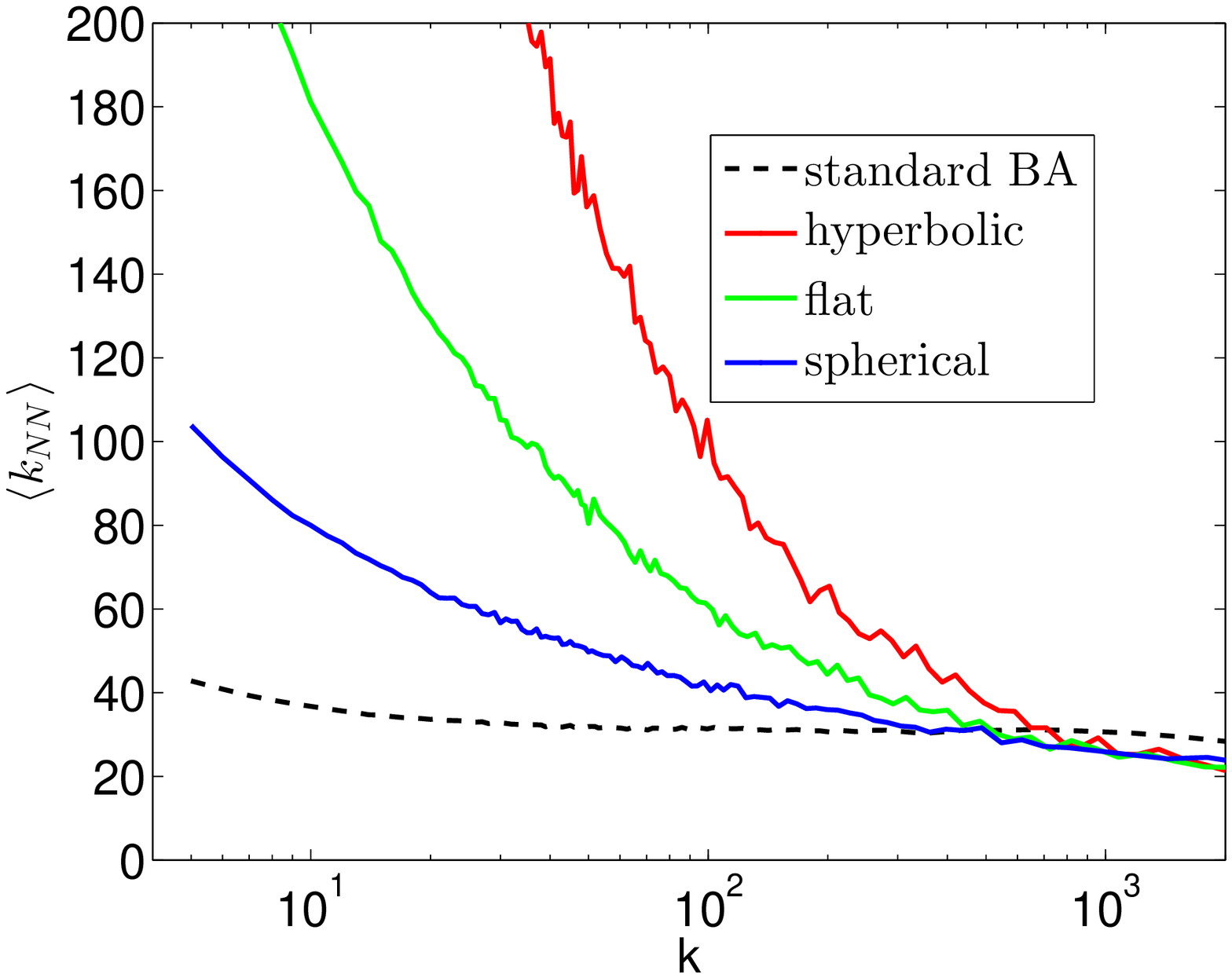}
\caption{Plot of average neighbour degree $\langle k_{NN}(k)\rangle$ (averaged over 50 simulations) for networks of size $2\cdot 10^5$ on disks of radius $r=1.5$ in two-dimensional spaces of curvature $+1,0,-1$ with connection function $\tilde{\sigma}(d)=\theta(1-d/r)$.}
\label{fig_theta} 
\end{figure}

As an example, we consider spatial networks with preferential attachment \cite{ferretti2011}. In these networks $S$ corresponds to a metric space and the connection function depends only on the distance $\sigma(x,y)=\tilde{\sigma}(d(x,y))$. We simulated network growth on disks in 2-dimensional spaces with constant curvature (sphere, flat and hyperbolic space) and calculated both the assortativity coefficient $\cass$ \cite{newman2003mixing} and the average nearest neighbour degree $\langle k_{NN}(k)\rangle$. We show the results in table \ref{tab_ass} and figures \ref{fig_exp16}, \ref{fig_exp4} and \ref{fig_theta}. 

It is apparent that the short-range connections and the spatial structure make the network slightly  more disassortative, because hubs tend to be sparse (close hubs compete between them and reduce their degree). Hyperbolic spaces at strong curvature show even higher disassortativity because of their almost star-like connectivity. 

Also, the generally low values of the assortativity coefficients suggest that they decrease with time as it happens in the BA model. 

\begin{table}[h]
\begin{center}
\begin{tabular}{|c|c|c|c|c|c|c|}
\multicolumn{7}{c}{$\cass$ for spatial networks:} \\
\hline
Space & \multicolumn{2}{|c|}{ Spherical} 	 & Flat 	 & \multicolumn{3}{|c|}{Hyperbolic}  \\  \hline 

Curvature & $
2.25$	& $
1$	 &  $
0$	 & $
-1$	 & $
-2.25$ 	 & $
-16$ \\  \hline 

	$\tilde{\sigma}=e^{-16d/r}$	 & -0.0139	& -0.0142	& -0.0145	& -0.0147	& -0.0151	& -0.0187 \\
	$\tilde{\sigma}=e^{-4d/r}$	& -0.0144	& -0.0151	& -0.0155	& -0.0162	& -0.0169	& -0.0136 \\
	$\tilde{\sigma}=\theta\left(1-\frac{d}{r}\right)$	& -0.0160	& -0.0158	& -0.0152	& -0.0149	& -0.0149	& -0.0909 \\ \hline
\end{tabular}
\end{center}
\caption{Values of the assortativity coefficient $\cass$ (averaged over 50 simulations) for networks of size $2\cdot 10^5$ on disks of radius $r=1$ in different two-dimensional spaces and for different connection functions $\tilde{\sigma}(d)$. (Note that the only parameter of the space is actually the product of the curvature and the squared radius.) The corresponding assortativity coefficient for the Barab\'asi-Albert model is  $\cass^{BA}=-0.0112$.}\label{tab_ass}
\end{table}

\section{Generalizations}

The Barab\'asi-Albert model is based purely on addition of nodes and preferential attachment. Realistic models can include many other ingredients: addition of extra links, rewiring and removal of links, directed links, variable initial degree, attachment functions that are only asymptotically linear in $k$, etc. Some generalizations of the BA model are reviewed in \cite{albert2002statistical,newman}.

In this section we present several variations on the heterogeneous models with preferential attachment presented in section \ref{sectionmodel}. For most of these generalizations, the degree distribution is a sum of power-laws, showing that scale-free or multi-scaling behaviour is a robust feature of preferential attachment models. The results are generally similar when different generalizations are combined together. 

The consistency equations for $ q(h)$ in these models 
can be obtained through a rate equation approach or, more easily, by using the continuum approach, i.e. the deterministic evolution of node degree $dk_i/dt=q(h_i)k_i/t$ or equivalently $k_i(t)=k_i(t_0)(t/t_0)^{q(h_i)}$ \cite{Barabasi1999mean}. The corresponding equations for $q(h)$ are exact, as explained in appendix \ref{stochastic}.

\subsection{Heterogeneity in initial degree}

We consider a model of growing networks with the usual preferential attachment rule (\ref{prefattachrule}). However, new nodes have a initial degree $m(h)$ that depends on their \hv\ $h$. 
The continuum equation for the node degree is
\beq
\frac{dk_i}{dt}=\sum_h p(h) m(h) \frac{\pr(h_i,h)k_{i}}{\sum_{l}
\pr(h_l,h)k_{l}}
\eeq
The degree distribution is given by equation
\beq
n_k=\sum_{h\in\Spa}\frac{\pro(h)}{q(h)m(h)}\left(\frac{k}{m(h)}\right)^{-(1+q(h)^{-1})} \theta(k-m(h)) \label{pkm}
\eeq
with $q(h)$ satisfying the consistency equations
\beq
q(h)=\sum_{l\in\Spa}  \frac{\pr(h,l) \pro(l)m(l)}{C(l)}
\quad,
\quad
C(h)=\sum_{l\in\Spa} \frac{\pr(l,h)\pro(l)m(l)}{1-q(l)} 
\eeq

Interestingly, if $\pr(h,h')=1$, the model is equivalent to a variation on the BA model with a random initial degree for the new nodes. In practice, the \hv\ is the initial degree $m$ of each node. Assume that the average initial connectivity $\bar{m}$ is finite. Then if the distribution of $m$ decays faster than $m^{-3}$, the degree distribution of this model for $k\gg \bar{m}$ is 
the same as the BA model. Instead, if the distribution of $m$ decays as a power law with exponent $-\alpha$ greater than $-3$, the sum in equation (\ref{pkm}) gives an additional factor $k^{3-\alpha}$ and therefore $n_k\sim k^{-\alpha}$. More generally, we can consider a general variation on the BA model with degree distribution $p_0(k)\sim k^{-\gamma}$ for fixed $m$, and modify this model to allow for a stochastic initial degree with distribution $p(m)\sim m^{-\alpha}$. In this case the degree distribution from equation (\ref{pkm}) is $n_k\sim k^{-\min(\alpha,\gamma)}$, in agreement with the formal results in \cite{deijfen2009preferential}.

\subsection{Heterogeneous links}
In this model, the attractiveness of a node does not depend only on the number of links, but also on the types of nodes to which they are attached. For examples, in copying vertex models or walking models \cite{vazquez2001disordered}, new nodes could preferentially explore existing nodes with a specific feature (e.g. search engines in the WWW example), therefore the effective preferential attachment dynamics would give an higher weight to the links that connect to this feature. 

The preferential attachment probability is a positive linear combination of $k_{i(h)}$, which is the number of links between the $i$th node and nodes with variable $h$:
\beq
\Pi(i)=\frac{\sum_{h'}\pr(h_i,h;h')k_{i(h')}}{\sum_{l}
\sum_{h'}\pr(h_l,h;h')k_{l(h')}}
\eeq
The distribution follows equation (\ref{distrk}) with quality $q(h)$ defined by the set of consistency equations
\begin{align}
q(h_i) & =\sum_{h,h'} \frac{\pr(h_i,h;h')f^{IN}(h_i,h')p(h)}{C(h)}\\
C(h) & =\sum_{h',h''}\pr(h'',h;h')p(h'')\cdot \\ 
\cdot & \left[f^{OUT}(h'',h') +  f^{IN}(h'',h')\frac{q(h'')}{1-q(h'')}\right] \nonumber \\
f^{OUT}(h_i,h')& = \sum_{h''}\frac{\pr(h',h_i;h'')p(h')}{C(h_i)} \cdot \\ 
\cdot & \left[f^{OUT}(h',h'')+ f^{IN}(h',h'')\frac{q(h')}{1-q(h')}\right] \nonumber \\
f^{IN}(h_i,h') & =\sum_{h''}\frac{p(h'')\sigma(h_i,h';h'')f^{IN}(h_i,h'')}{q(h_i)C(h')}
\end{align}


\subsection{Shifted preferential attachment}
The preferential attachment rule is modified by the addition of a positive term $a(h_i,h)$ independent of the degree but dependent on the \hvs. Similar models without features were proposed in \cite{krapivsky2001organization,dorogovtsev2000structure}.
\beq
\Pi(i)=\frac{\pr(h_i,h)k_{i}+a(h_i,h)}{\sum_{l}
\pr(h_l,h)k_{l}+a(h_l,h)}
\eeq
The degree distribution for large $k$ follows equation (\ref{distrk}) with quality $q(h)$ defined by (\ref{qdef}) and $C(h)$ defined by 
\beq
C(h)=\sum_{h'}\left[\frac{\pr(h',h)}{1-q(h')}+\frac{a(h',h)}{m}\right]p(h')
\eeq

\subsection{Directed links}
In this model the preferential attachment probability is proportional to the number of incoming links $k^{IN}$:
\beq
\Pi(i)=\frac{\pr(h_i,h)(k^{IN}_{i}+a(h_i))}{\sum_{l}
\pr(h_l,h)(k^{IN}_{l}+a(h_l))}
\eeq
Note that the positive term $a(h)$ is needed to specify the initial attachment probability because initially $k^{IN}=0$. The results are similar to the shifted preferential attachment case, with $C(h)$ defined by
\beq
C(h)=\sum_{h'}\pr(h',h)\left[\frac{q(h')}{1-q(h')}+\frac{a(h)}{m}\right]p(h')
\eeq

\subsection{Addition of links}
In this model, in addition to the usual growth rules, extra links are added at rate $mr_+$ and attached to nodes $i,j$ according to the probability
\beq
\Pi_+(i,j)=\frac{\pr_+(h_i,h_j)k_{i}k_{j}}{\sum_{r,s}
\pr_+(h_r,h_s)k_{r}k_{s}}
\eeq
This model can be solved similarly to the usual one under some extra assumptions, like the scaling
\beq
\sum_{r,s}
\pr_+(h_r,h_s)k_{r}k_{s}=m^2C_+t^2+o(t^2)
\eeq
The degree distribution follows equation (\ref{distrk}) with quality $q(h)$ defined by
\begin{align}
q(h_i)=&\sum_h  \left[\frac{\pr(h_i,h)}{C(h)}+\frac{2r_+\sigma_+(h_i,h)}{C_+(1-q(h))}\right]p(h) \\
C(h)=& \sum_{h'}\frac{\pr(h',h)p(h')}{1-q(h')}\\
C_+=& \sum_{h,h'}\frac{\sigma_+(h',h)p(h)p(h')}{(1-q(h))(1-q(h'))}
\end{align}

\subsection{Preferential rewiring in directed networks}

Modifications to the usual preferential attachment growth of the node degree include also possible losses of links, either because of removal or rewiring \cite{albert2000topology}. An high rate of link removal/rewiring could result in a degree distribution with an exponential tail instead of the usual power-law tail. However, if the removal/rewiring process is not too fast, the resulting distribution is typically a power-law with an exponent dependent on the rates of the different processes.
 
Several heterogeneous models with preferential rewiring and/or removal of links can be analyzed with the techniques of this paper
.  In these models the growth of node degrees can be characterized by an effective fitness $q(h)$ and the stochastic noise due to link addition and removal does not change significantly the tail of the degree distribution, as explained in appendix \ref{stochastic}. Here present a simple example of such a model. 

We consider a directed network growing under the same rules as the model (\ref{prefattachrule}) and with rewiring taking place at rate $mr$. In each rewiring process, a random link is selected with probability proportional to $\sigma_-(h_{in},h_{out})$ where $h_{in}$ and $h_{out}$ are the variables of the attached nodes. The ingoing end of the link is then detached and reattached to another node according to the probability
\beq
\Pi_+(i)=\frac{\sigma_+(h_i,h_{out})k_i^{IN}}{\sum_j\sigma_+(h_j,h_{out})k_j^{IN}}
\eeq 

In this model, at least for large $k$, the degree distribution follows a multi-scaling behaviour similar to equation (\ref{distrk}), given by 
\beq
n_k\simeq\sum_{h\in\Spa}\theta(q(h))\frac{\pro(h)}{q(h)m}\left(\frac{k}{m}\right)^{-(1+q(h)^{-1})}  \label{distrkrewiring}
\eeq 
with quality $q(h)$ defined by the set of consistency equations
\begin{align}
q(h_i)=&\sum_h \frac{\pr(h_i,h)p(h)}{C(h)}+\frac{r\Sigma_+(h_i)}{C_+}+\\
&-\sum_h\frac{r\sigma_-(h_i,h)f^{IN}(h_i,h)}{C_-} \quad , \nonumber \\
C(h)=& \sum_{h'}\frac{\pr(h',h)p(h')}{1-q(h')}\\
C_+=& \sum_{h}\frac{\Sigma_+(h)q(h)p(h)}{1-q(h)} \\
C_-= & \sum_{h,h'}\frac{\sigma_-(h',h)f^{IN}(h',h)q(h')p(h')}{1-q(h')}& \\
f^{IN}(h_i,h)=& \frac{\pr(h_i,h)p(h)/C(h)+r\Sigma_+(h_i)/C_+}{q(h_i)+r\sigma_-(h_i,h)/C_-}\\
\Sigma_+ (h_i)= &\sum_{h,h'} \frac{\sigma_+(h_i,h)\sigma_-(h',h)f^{IN}(h',h)q(h')p(h')}{
C_-(1-q(h'))}
\end{align}
In this model (and more generally in models including rewiring/removal of links) the quality $q(h)$ can also be negative, thus requiring the factor $\theta(q(h))$ in equation  (\ref{distrkrewiring}).

\subsection{Fixing the connection probability between \hvs}

The last variation is a more radical departure from the heterogeneous models presented in this paper, since it is a modification of the attachment probability (\ref{prefattachrule}). In this model, a new node with \hv\ $h$ attaches to nodes with different variables $h'$ according to a probability $\pi(h'|h)$ independent of the degrees. Then, once a \hv\ $h'$ is chosen at random according to $\pi(h'|h)$, a specific node $i$ with $h_i=h'$ is chosen according to the usual preferential attachment rule. (If no such node exists, another variable $h''$ is chosen according to $\pi(h''|h)$ .) 
Asymptotically, the overall probability is then
\beq
\Pi(i)=\pi(h_i|h)\frac{k_i}{\sum_k k N_k(h_i,t)}\label{modprefattachrule}
\eeq
It is possible to define a quality $q(h)$ also for these models. Assuming the scaling
\beq
\sum_{\{j|h_j=h\}}k_j = mp(h)C(h)t+o(t)
\eeq
we obtain $C(h)=1/(1-q(h))$ and the quality can be obtained explicitly:
\beq
q(h_i)=\frac{\sum_{h}\pi(h_i|h)p(h)}{p(h_i)+\sum_{h}\pi(h_i|h)p(h)}\label{qeqmod}
\eeq
The degree distribution follows the usual equation (\ref{distrk}). This model works if the \hs\ $\Spa$ is discrete and finite, but can be generalized to continuous spaces through discretization and equation (\ref{qeqmod}) is valid with $p(h)$ and $\pi(h'|h)$ interpreted as probability densities.

\section{Conclusions}

The addition of node features to  growing network models with preferential attachment  is an important step towards realistic network modeling and results in a wide class of models, for which this paper provides several analytical results. In particular, this work shows that the power-law scaling of the degree distribution generated by preferential attachment is quite robust with respect to the heterogeneity between nodes. The main effect of heterogeneity is the emergence of an ``effective fitness'' $q(h)$ for each class of nodes, therefore their degree distribution resembles the fitness model of Bianconi and Barab\'asi \cite{bianconi2001competition,bianconi2001bose}.

Beyond the degree distribution, other network properties were studied. The clustering coefficient of these networks disappears for large network size, a property shared with the BA model. Negative degree correlations are also present in these models, along with non-trivial mixing patterns among features. Both small clustering coefficients and disassortative mixing  are therefore outcomes of the preferential attachment mechanism in general growing networks.      


The effect of the \hvs\ $h$ associated to each node has been presented as non-random, but the formalism applies to any kind of heterogeneity. In particular it is easy to include random variables or \hvs\ with random effects as well, as long as their values do not change with time. In fact, any random effect can be parametrized by some extra random variables $\chi$ with a distribution $p_R(\chi)$. Then it is possible to redefine a non-random variable $\tilde{h}=(h,\chi)$ with frequency $\tilde{p}(\tilde{h})=p(h)p_R(\chi)$. The connection function $\tilde{\sigma}(\tilde{h}, \tilde{h}')$ now takes into account both random and non-random components. So the formalism captures stochastic as well as deterministic node features.

Moreover, the growth of many  scale-free networks is  based  on some  local dynamics such as the  vertex copying/duplication rules for growth of  molecular networks. However, from the point of view of the link distribution, the local dynamics often results in an effective preferential attachment mechanism. Our methods and results on degree distribution and assortativity  apply to these models as well, if we take the connection probability (\ref{prefattachrule}) as an effective dynamics for the growth of the node connectivities. 
Therefore the main results of this paper, i.e. power-law multiscaling of the degree distribution and disassortative mixing in degree, are generally valid for models with effective preferential attachment. 
On the other way, the clustering coefficient depends on the specific model and not only on the effective form (\ref{prefattachrule}) for the attachment probability, therefore our proof that the clustering vanishes in the thermodynamic limit is valid only for models with pure preferential attachment.

In future works it would be very interesting to map the metric space implied by the class of growing network models discussed in this paper  with the  hidden metrics recently introduced to model complex networks in hyperbolic spaces \cite{krioukov2009curvature}.
Finally, an interesting extension of the model would be to include \hvs\ that fluctuate in time, in order to  determine how time-dependent heterogeneities affect the power-law behaviour of the degree distribution  and the other properties discussed here, and features that coevolve with the network, for example spaces that expand with time while the node density remains constant.

\begin{acknowledgments}
We thank M. Bogu\~na and M. Mamino for useful discussions. L.F. acknowledges support from CSIC (Spain) under the JAE-doc program.
\end{acknowledgments}

\bibliographystyle{apsrev4-1.bst}
\bibliography{network2_prefattach}

\appendix

\section{Clustering in the Barab\'asi-Albert model and in heterogenous models}\label{clustba}
In a general heterogenous model, we consider a triplet of nodes born at times $t_A<t_B<t_C$ with qualities $q_A,q_B,q_C$. The average number of triangles can be found by integrating the probability that all three nodes are connected on the birth times $t_A,t_B,t_C$:
\begin{align}
n_\mathrm{triangles}&=\int_1^\N dt_C\int_1^{t_C}dt_A\int_{t_A}^{t_C}dt_B\frac{\sigma(h_A,h_C)m(t_C/t_A)^{q_A}}{mC(h_C)t_C}\cdot \\
\cdot &\frac{\sigma(h_B,h_C)(m-1)(t_C/t_B)^{q_B}}{mC(h_C)t_C}\frac{\sigma(h_A, h_B)m(t_B/t_A)^{q_A}}{mC(h_B)t_B} \cdot \nonumber 
\end{align}

The average number of triples can be found in a similar way. These networks can contain three kind of triples: $A\leftarrow B\leftarrow C$, $A\leftarrow C\rightarrow B$ and $B\rightarrow A\leftarrow C$. Since each new node has $m$ outgoing links, it increases the number of triples $A\leftarrow C\rightarrow B$ by a factor $m(m-1)/2$ and the number of triples $A\leftarrow B\leftarrow C$ by a factor $m^2$, independently of the model, so their total number is 
$m(3m-1)\N/2$. 
The number of triples $B\rightarrow A\leftarrow C$ is given by the integral
\begin{align}
&n_{\mathrm{triples}\ B\rightarrow A\leftarrow C}=\int_1^\N dt_C\int_1^{t_C}dt_A\int_{t_A}^{t_C}dt_B\\
& 
\frac{\sigma(h_A,h_C)m(t_C/t_A)^{q_A}}{mC(h_C)t_C}\frac{\sigma(h_A,h_B)m(t_B/t_A)^{q_A}}{mC(h_B)t_B}\nonumber
\end{align}

We give the full result for BA networks, since the other cases are straightforward (but cumbersome) generalizations. In the BA case we have $q_A=q_B=q_C=1/2$ and the clustering coefficient is
\beq
C^{BA}_{clust}=\frac{m(m-1)\ln^3 \N}{8\left(m\N\ln \N-\N+m\N^{1/2}\right)}
\eeq
including some finite corrections to the leading behaviour $C(\N)\propto\ln^2\N/\N$.


\section{Assortativity in heterogeneous models}\label{analyticassort}
 
Following \cite{krapivsky2001organization}, the rate equation for $N_{k,l}^{(h,i)}=n_{k,l}^{(h,i)}\cdot mt$ is 
\begin{align}
\frac{dN^{(h,i)}_{k,l}}{dt}=&\frac{q(h)}{t}\left[(k-1)N^{(h,i)}_{k-1,l}-kN^{(h,i)}_{k,l}\right]+ \label{eqassrate}\\
+&\frac{q(i)}{t}  \left[(l-1)N^{(h,i)}_{k,l-1}
-lN^{(h,i)}_{k,l}\right]+ \nonumber \\
+ &\delta_{l,m}\frac{\sigma(h,i)p(i)}{C(i)t}(k-1)N_{k-1}(h,t)\nonumber
\end{align}
which is equivalent to equation (\ref{eqass}) after substituting $n_{k,l}^{(h,i)}$ and rearranging. 
The solution can be found in terms of the generating function 
\beq
g(x,y)=\sum_{k=m+1}^\infty\sum_{l=m}^\infty x^k y^l n_{k,l}^{(h,i)}
\eeq
and the result is
\begin{align}
g(x,y)&=\frac{\sigma(h,i)p(i)}{C(i)q(h)}\frac{(1-x)^{(1+mq(i))/q(h)} y^m}{x^{1/q(h)}}\cdot \nonumber\\
\cdot & \int_0^x\frac{dr\ f(r)r^{\frac{1+mq(i)-q(h)}{q(h)}}(1-r)^{-1-\frac{1}{q(h)}}}{\left[(1-x)^{\frac{q(i)}{q(h)}}y r^{\frac{q(i)}{q(h)}} + x^{\frac{q(i)}{q(h)}}(1-y) (1-r)^{\frac{q(i)}{q(h)}} \right]^m} \label{gfass}
\end{align}
where $f(r)$ is the generating function of $(k-1)n_{k-1}(h)$:
\begin{align}
f(r)&=\frac{p(h)\Gamma(m+q(h)^{-1})}{q(h)\Gamma(m)}\left(\frac{1-x}{x}\right)^{q(h)^{-1}-1}\cdot \nonumber \\ \cdot &\int_0^x dz\ z^{m-1}\left(\frac{z}{1-z}\right)^{q(h)^{-1}}
\end{align}
For $m=1$ the generating function (\ref{gfass}) can be expanded into the solution (\ref{gensolass}).

For $m=1$ it is easy to expand $g(x,y)$ in power of $y$ and then apply Tauberian theorems to the functions $l!(\partial/\partial y)^l g(x,y)|_{y=0}$, obtaining the scaling $n_{k,l}^{(h,i)}\sim k^{-1/q(h)}l^{-1-q(h)/q(i)}$ for fixed $l$ and large $k$.

Since it is not easy to further understand the behaviour of $n_{k,l}^{(h,i)}$ from the above solution, we approximate the differences in square brackets in (\ref{eqassrate}) as derivatives $-\partial (kN^{(h,i)}_{k,l})/\partial k$ and $-\partial (lN^{(h,i)}_{k,l})/\partial l$, rearrange and obtain the differential equation $n^{(h,i)}_{k,l}=q(h)\partial (kn^{(h,i)}_{k,l})/\partial k+q(i)\partial (l n^{(h,i)}_{k,l})/\partial l$. Its solutions have the form
\beq
n^{(h,i)}_{k,l}\sim f(k^{1/q(h)}l^{-1/q(i)})k^{-1-1/2q(h)}l^{-1-1/2q(i)}\label{eqassf}
\eeq 
where $f$ is an arbitrary differentiable function. 

Now we obtain the scaling of the exact boundary conditions $n^{(h,i)}_{k,m}$ and $n^{(h,i)}_{m+1,l}$. The first satisfies the equation
\begin{align}
n^{(h,i)}_{k,m}(1+q(h)k+q(i)m)&=q(h)(k-1)n^{(h,i)}_{k-1,m}\\
&+\frac{\sigma(h,i)p(i)}{C(i)}(k-1)n_{k-1}(h)\label{eqassb1}\nonumber
\end{align}
Its solution is
\begin{align}
n^{(h,i)}_{k,m}=&\frac{\sigma(h,i)p(i)p(h)}{C(i)q(h)^2}\frac{\Gamma(m+q(h)^{-1})\Gamma(k)}{\Gamma(m)\Gamma(k+1+(1+mq(i))/q(h))}\cdot \nonumber \\ &  \cdot \sum_{j=m+1}^k\frac{\Gamma(j+(1+mq(i))/q(h))}{\Gamma(j+q(h)^{-1})} \sim k^{-1/q(h)}
\end{align}
Matching the scaling in $k$ with equation (\ref{eqassf}) gives $f(k^{1/q(h)})k^{-1-1/2q(h)}\sim k^{-1/q(h)}$, so $f(x)\sim x^{q(h)-1/2}$ and therefore $n^{(h,i)}_{k,l}\sim k^{-1/q(h)} l^{-1-q(h)/q(i)}$ for $l\ll k$, consistent with exact results. 

The second boundary $n^{(h,i)}_{m+1,l}$ satisfies the equation
\begin{align}
n^{(h,i)}_{m+1,l}&(1+q(h)(m+1)+q(i)l)=\\= &q(i)(l-1)n^{(h,i)}_{m+1,l-1}
+\delta_{l,m}\frac{\sigma(h,i)p(i)}{C(i)}mn_{m}(h)\nonumber
\label{eqassb2}
\end{align}

Its solution is
\begin{align}
n^{(h,i)}_{m+1,l}&=
\frac{\sigma(h,i)p(i)p(h)m}{C(i)(m+q(h)^{-1})q(h)q(i)} \cdot\\
&\cdot \frac{\Gamma(l)\Gamma(m+(1+(m+1)q(h))/q(i))}{\Gamma(m)\Gamma(l+1+(1+(m+1)q(h))/q(i))} \nonumber\\
&\sim l^{-1-(1+(m+1)q(h))/q(i)}\nonumber
\end{align}
so the matching $f(l^{-1/q(i)})l^{-1-1/2q(i)}\sim l^{-1-(1+(m+1)q(h))/q(i)}$ gives $f(x)\sim x^{1/2+(m+1)q(h)}$ and therefore $n^{(h,i)}_{k,l}\sim k^m l^{-1-(1+(m+1)q(h))/q(i)}$ for $k\ll l$. 

For the BA model (i.e. $q(h)=q(i)=1/2$) with $m=1$, we obtain $n^{(h,i)}_{k,l}\sim k^{-2} l^{-2}$ for $k\gg l$ and $k l^{-5}$ for $k\ll l$, in agreement with exact results \cite{krapivsky2001organization}.

\section{Stochastic effects and removal of links}\label{stochastic}

Some of the models presented in this paper involve with not only addition of new links, but also removal of existing links. For example, the rewiring process is equivalent to removing and then adding a link to the network.

In these models, the continuum equation for the degree of a node \cite{Barabasi1999mean} is simply ${dk}/{dt}=(q_+ - q_-){k}/{t}$, where $q_+$ and $q_-$ represent asymptotic growth and reduction coefficients coming from addition and deletion of links, respectively. The birth rate of nodes is constant, therefore the continuum approach predicts that the degree distribution is the usual power law $n_k\propto k^{-1-1/(q_+ - q_-)}$ expected for growing networks with preferential attachment. However, the continuum equation could be wrong in predicting the degree distribution or the consistency equations in these models. In this appendix we examine these aspect in more detail.

First, the consistency equations have typically the form 
\begin{align}
f(h)t=&\left\langle \sum_i g(h_i)k_i(t) \right\rangle \quad \\
&\Updownarrow \nonumber\\
\quad f(h)=\sum_h g(h) p(h) \cdot & \frac{1}{p(h)t}\sum_{\{i|h_i=h\}}\left\langle k_i(t) \right\rangle \nonumber
\end{align} 
where $\langle \ldots \rangle$ represents an average over realizations of the process. Therefore these equations depend only on the average degree $\left\langle k \right\rangle$ of the nodes. But the asymptotic evolution of the average degree $\left\langle k(t) \right\rangle$ is described precisely by the continuum equation for $k(t)$, therefore the consistency equations predicted by the continuum approach are exact. For example, equation (\ref{ceq}) can be obtained equivalently from the rate equation approach or the continuum approach, i.e., from the first or the second line of equation (\ref{nkeq}).

Second, while $q_+$ and $q_-$ are positive by definition, there is no reason for $q_+-q_-$ to be positive definite. In fact, there are models where all nodes have negative global quality $q_+-q_-<0$. The degree distribution of these models deviates significantly from the scale-free behaviour, since on average the connectivity of all nodes decreases with time and nodes of high degree can appear only due to the effect of fluctuations, therefore the tail of the distribution falls exponentially. (In particular, applying the Langevin equation approximation presented below, it should fall faster than $e^{-\sqrt{k}}$.)  
The exact degree distribution for these models will be presented elsewhere. 

However, in the present paper we focus on models where at least some classes of nodes have positive quality $q_+-q_->0$. In these models there are two kinds of nodes: those with $q_+>q_-$ (and therefore positive average growth rate), for which the continuum approach could work, and those with $q_+<q_-$ (and negative average growth rate), for which the continuum approach fails as described above. However, the tail of the latter nodes falls rapidly and do not contribute to the degree distribution, so it is sufficient to calculate the degree distribution using the continuum approximation for the nodes with $q_+-q_->0$ and discard the others.




Third, even for nodes with $q_+-q_->0$, rewiring and removal enhance the stochasticity of the process, potentially affecting the distribution. To take this noise into account, we promote the above equation to a  Langevin equation obtained by adding a stochastic term $+\sqrt{(q_+ + q_-)k/t}\cdot\eta(t)$ with $\eta(t)$ a white Gaussian noise with variance 1. Defining $q=q_+-q_-$ and $Q=q_++q_-$, we finally obtain the integrated Fokker-Planck equation 
\beq
\frac{\partial n(k,t)}{\partial t}=-\frac{n(k,t)}{t}-\frac{q}{t}\frac{\partial (k n(k,t))}{\partial k}+\frac{Q}{2t}\frac{\partial^2 (kn(k,t))}{\partial k^2}
\eeq
for the average degree distribution $n(k,t)=\int_1^tdt_0P(k,t|t_0)/t$, where $P(k,t|t_0)$ is the distribution for a node born at time $t_0$. The rate equation approach leads to the same equation. The (normalizable) stationary solution of this equation can be derived as a power series:
\beq
n_k\propto k^{-1-q^{-1}}\left[1+\sum_{n=1}^\infty (1+nq)\frac{\Gamma(n+q^{-1})^2}{\Gamma(n+1)}\left(-\frac{Q}{q}\right)^n k^{-n}\right]\label{pkborel}
\eeq
This is an asymptotic series that can be resummed by Borel  summation in the variable $z=k^{-1}$. The leading contribution to the Borel transform of the sum in (\ref{pkborel}) behaves as $\mathcal{B}(z)\sim Qz$ for $z\rightarrow 0$, therefore for large $k$ the Borel sum is $Qk^{-1}$ and the resulting degree distribution is $n_k\sim k^{-1-q^{-1}}(1+O(Qk^{-1}))$, thereby confirming the validity of the simple continuum approach (without the diffusion term) for $k\gg Q$.

\end{document}